\documentclass{jpsj3}
\usepackage{txfonts}
\usepackage{widetext}
\usepackage{overcite}
\usepackage{graphicx}

\title{Backward Simulation of Stochastic Process \\ 
Using a Time Reverse Monte Carlo Method}

\author{Shinichi Takayanagi$^1$\thanks{stakaya@ism.ac.jp} and Yukito Iba$^2$\thanks{iba@ism.ac.jp}}
\inst{$^1$SOKENDAI, 10-3 Midori-cho, Tachikawa, Tokyo 190-8562, Japan \\
$^2$The Institute of Statistical Mathematics, 10-3 Midori-cho, Tachikawa, Tokyo 190-8562, Japan \\
} 

\abst{
The ``backward simulation'' of a stochastic process is defined as the stochastic dynamics that trace a time-reversed path from the target region to the initial configuration. If the probabilities calculated by the original simulation are easily restored from those obtained by backward dynamics, we can use it as a computational tool. It is shown that the na\"{i}ve approach to backward simulation does not work as expected. As a remedy, the time reverse Monte Carlo method (TRMC) based on the ideas of sequential importance sampling (SIS) and sequential Monte Carlo (SMC) is proposed and successfully tested with a stochastic typhoon model and the Lorenz 96 model. TRMC with SMC, which contains resampling steps, is found to be more efficient for simulations with a larger number of time steps. A limitation of TRMC and its relation to the Bayes formula are also discussed.
}

\begin{document}
\maketitle

\section{Introduction}

\begin{figure*}[tbp]
	\begin{tabular}{c}
		\begin{minipage}{60mm}
			\centering
			\includegraphics[width=55mm]{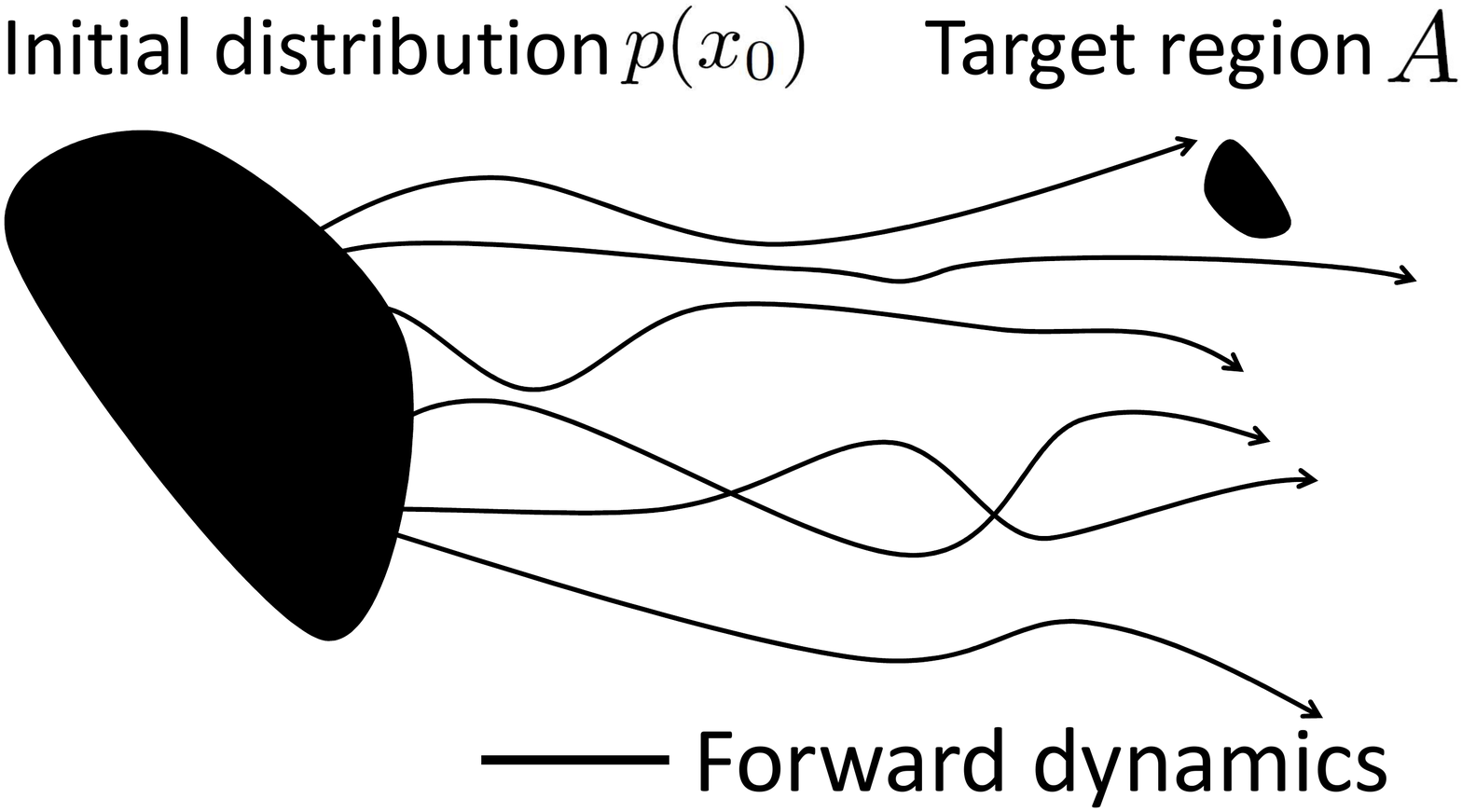}
		\end{minipage}
		\begin{minipage}{60mm}
			\centering
			\includegraphics[width=55mm]{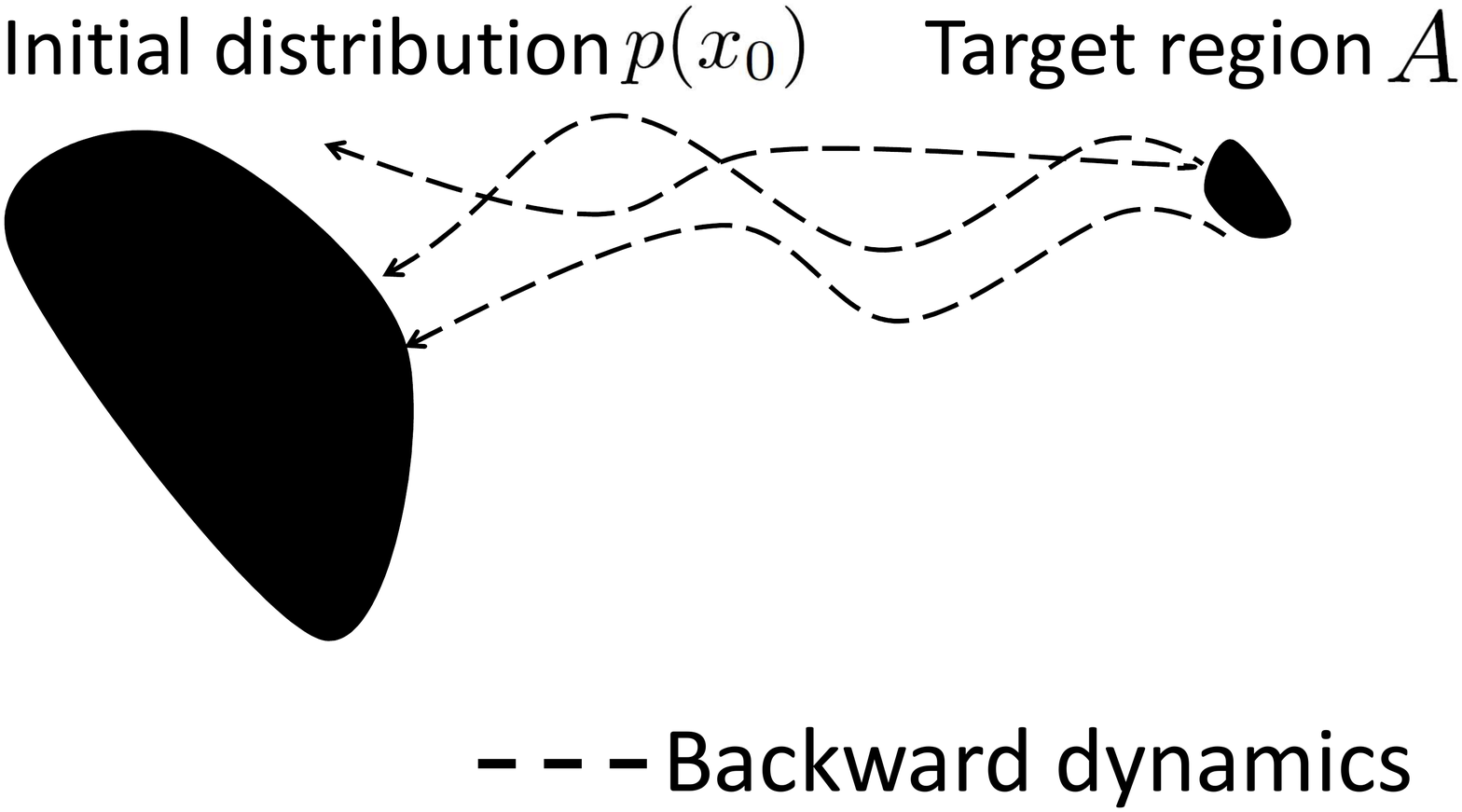}
		\end{minipage}
	\end{tabular}
	\caption{
		Forward and backward simulations.
		The forward simulation is inefficient when the target region $A$ is much smaller than the support of the initial distribution $p(x_0)$.
		The backward simulation simulates paths from the target region $A$ to the support of the initial distribution $p(x_0)$.	
	}
	\label{fig:concept}
\end{figure*}

In this paper, we discuss ``backward simulation,'' which traces a time-reversed path from a target region $A$ to the initial configuration (Fig.~\ref{fig:concept}). If the outputs of the original simulation (``forward simulation'') are easily restored from those obtained by backward dynamics, we can use backward simulation as a computational tool. In particular, the time required to calculate the probability to reach $A$ from the initial configurations can be significantly reduced when the target region $A$ is small but the initial distribution is broad.
An example is a computation of the probability that a typhoon will hit the Tokyo area exactly under a given stochastic model (Sect.~\ref{subsec:stochastic_typhoon}).

It is, however, difficult to design backward dynamics with the desired properties. Specifically, consider the forward dynamics of a $D$-dimensional stochastic process $X$ defined by 
\begin{equation}
X_{i+1} = g(X_i) + \eta_i,
\label{eq:forward_general}
\end{equation} 
where $\eta_i$ is an independent noise that obeys an arbitrary distribution and the function $g: \mathbb{R}^{D} \rightarrow \mathbb{R}^{D}$ describes noiseless forward dynamics.
Then, a na\"{i}ve way to derive a time-reversed equation is 
to rearrange Eq.~\eqref{eq:forward_general} as
\begin{equation}
X_i = g^{-1}(X_{i+1} - \eta_i).
\label{eq:forward_inverse_general}
\end{equation} 
Here, we assume that function $g$ is a one-to-one and onto function and
denotes the inverse function of $g$ as $g^{-1}$. We can construct a time-reversed path iteratively, using Eq.~\eqref{eq:forward_inverse_general} and the independent realization of $\eta_i$, starting from the target region. It defines an apparently natural candidate for backward dynamics.  

Surprisingly, this na\"{i}ve method does not work as expected; it does not reproduce the correct probabilities defined by the forward simulation, and the calculation of factors required to correct the bias is often computationally expensive. This becomes clear in Sect.~\ref{sec:fnb}. Furthermore, the computation of $g^{-1}$ in Eq.~\eqref{eq:forward_inverse_general} is time-consuming and reduces the efficiency of the computation.    

The aim of this research is to draw attention to these facts and propose an algorithm that partially resolves the problem. We named this algorithm the time reverse Monte Carlo method (TRMC). TRMC is based on the ideas of sequential importance sampling (SIS)~\cite{doucet2001introduction, liu2008monte} and sequential Monte Carlo (SMC)~\cite{doucet2001introduction,doucet2009tutorial, tailleur2007probing, zuckerman2010statistical}. We discuss TRMC based on SIS in Sects.~\ref{sec:trmc} and \ref{sec:isde} and its improved version based on SMC in Sect.~\ref{sec:is_resampling}. 
There have been several studies using ``path reweighting'' in computational chemistry~\cite{donati2017girsanov, chodera2011}.
These studies used reweighting for different purposes.

TRMC based on SIS is tested for a stochastic typhoon model and the Lorenz 96 model in Sect.~\ref{sec:er}. The improved version with SMC is also tested in Sect.~\ref{sec:is_resampling} for simulations with a larger number of steps. In these examples, TRMC provides unbiased estimates of the probabilities without expensive computation. In Sect.~\ref{sec:dis}, we discuss its relation to the Bayes formula, as well as the possible improvement and limitations of TRMC. 

Time-reversed dynamics itself was discussed in several studies~\cite{anderson1972factors, haussmann1986time, millet1989integration}, mostly from a theoretical viewpoint. On the other hand, related computational problems are  
found in data science, especially in time-series analysis using 
state-space models~\cite{briers2010smoothing, lindsten2013backward, isard1998smoothing}. Our problem can formally be regarded as a limiting case of the ``smoothing'' part of these algorithms, where only one observation (``target'') is available at the end of the time series. There are, however, important differences from our problem, which are discussed in Sect.~\ref{sec:dis}.
Studies related to the statistical inference 
on a discrete state stochastic process, such as  
genepropagation~\cite{liu2008monte, griffiths1994simulating} and 
information source detection~\cite{zhu2016information} were also reported.
These studies, however, did not consider 
dynamical systems of continuous variables.

\section{Failure of Na\"{i}ve Method} \label{sec:gs}\label{sec:fnb}

Here, we provide a detailed discussion of the na\"{i}ve method and its drawbacks, 
which form the motivation for our algorithm. 
Before providing details, we formulate the problem.
Let $S_{T} = \left\{0 = t_0 \le t_1 \cdots \le t_{N} = T\right\}$ be a partition of the interval $[0, T]$, and let step size $\Delta t= t_{i+1}-t_{i}$ be a constant; $x_i$ is used to represent the value of stochastic process $X$ at time point $t_i$. The transition probability density from $x_{i}$ to $x_{i+1}$ defined by Eq.~\eqref{eq:forward_general} is denoted as $p\left( x_{i+1} | x_{i} \right)$.
We consider an estimation of the probability 
\mbox{$P(X_N \in A)$}
that $X_N$ hits a small target region $A$ in the $D$-dimensional space.
The probability is formally written as
\begin{align}
	P(X_N \in A) 
	=
	\int dx_{0:N} 
	\boldsymbol{1}_{x_N \in A}
	\left\{ \prod_{i=0}^{N-1}  p\left( x_{i+1} | x_{i} \right) \right\}
	p(x_{0})
	,
	\label{eq:fp}
\end{align}
where $\boldsymbol{1}_{x \in A}$ is the indicator function that takes value 1 when $x \in A$, and 0 otherwise, and 
$p(x_0)$ is the initial distribution of the forward simulation.
Hereafter,  $dx_{k:l}$ indicates $dx_{k}dx_{k+1} \cdots dx_{l}$ for $k \le l$. 

A na\"{i}ve method is defined as a repeated simulation 
with a uniformly distributed initial condition in 
the target region $A$ using Eq.~\eqref{eq:forward_inverse_general}.
Initially, it appears sufficient to evaluate \mbox{$P(X_N \in A)$} as \mbox{$\frac{1}{M} \sum_{j=1}^{M}p(x_0^{(j)})$}.
However, there are two problems with this na\"{i}ve method.
First, the exact computation of $g^{-1}$ in Eq.~\eqref{eq:forward_inverse_general} is not easy.
Computing $g^{-1}$ using numerical root-finding techniques such as the Newton-Raphson method is computationally intensive and its severity increases as the dimension increases.

Second, this computation does not reproduce the correct probability \mbox{$P(X_N \in A)$} even with the exact $g^{-1}$.
To understand this problem, we show the difference between the forward simulation and the na\"{i}ve method.
Let us define 
\begin{equation}
Y_{i}=X_{i}-\eta_{i-1} = g(X_{i-1}); i \in [1, \dots, N].
\label{eq:y_definition}
\end{equation}
Using this definition, we can rewrite Eq.~\eqref {eq:forward_inverse_general} as 
\begin{align}
	Y_{i} + \eta_{i-1} = g^{-1}(Y_{i+1}) .
	\label{eq:be_y}
\end{align}

Equation \eqref{eq:be_y} can be simplified into
\begin{align}
	Y_{i} =  g^{-1}(Y_{i+1}) - \eta_{i-1}.
	\label{eq:backward_naive}
\end{align}

The probability calculated using Eq.~\eqref{eq:backward_naive} corresponds to  the equation
\begin{align}
	\int dy_{1:N} dx_{N}
	\boldsymbol{1}_{x_N \in A}
	\tilde{p}_{f} \left( y_{N} | x_{N} \right)
	\left\{
	\prod_{i=1}^{N-1}
	\tilde{p}\left( y_{i} | y_{i+1} \right)
	\right\}
	p(g^{-1}(y_1))
	,
	\label{eq:bp_naive}	
\end{align}
where $\tilde{p}_{f} \left( y_N | x_N \right)$ is the transition probability density from $x_N$ to $y_N$ defined by Eq.~\eqref{eq:y_definition} with $i=N$ and
$\tilde{p}\left( y_{i} | y_{i+1} \right)$ is the transition probability density from $y_{i+1}$ to $y_{i}$ defined by Eq.~\eqref{eq:backward_naive}.
An initial condition $x_N$ is uniformly distributed in the target region $A$.

We have to introduce the Jacobian of function $g$ so that Eq.~\eqref{eq:bp_naive} is consistent with Eq.~\eqref{eq:fp}.
To show this, Eq.~\eqref{eq:fp} is rewritten using equations
\begin{align}
	p(x_{i}|x_{i-1})
	dx_{i}
	&=
	\left|\mathrm{det}(J_{g^{-1}}(y_{i+1}))\right|
	\tilde{p}\left( y_{i} | y_{i+1} \right)
	dy_i, \\
	i &\in [1,\cdots, N-1] \nonumber \\
	p(x_{0})
	dx_{0}
	&=
	\left|\mathrm{det}(J_{g^{-1}}(y_{1}))\right|
	p\left( g^{-1}(y_{1}) \right)
	dy_1,	
\end{align}
where $\left|\mathrm{det}(J_{g^{-1}}(y_{i}))\right|$ is the absolute value of the Jacobian of function $g^{-1}$.
As a result, the probability $P(X_N \in A)$ is calculated as
\begin{widetext}
	\begin{align}
		P(X_N \in A) 
		&=
		\int dy_{1:N}
		dx_N
		\boldsymbol{1}_{x_N \in A}
		J(y_1, \dots, y_N)
		\tilde{p}_{f}\left( y_{N} | x_{N} \right)
		\left\{
		\prod_{i=1}^{N-1}
		\tilde{p}\left( y_{i} | y_{i+1} \right)
		\right\}
		p\left( g^{-1}(y_{1}) \right)
		, 
		\label{eq:bp_naive_correct}	
		\\
		J(y_1, \dots, y_N)
		&=
		\left\{
		\prod_{i=0}^{N-1}
		\left|\mathrm{det}(J_{g^{-1}}(y_{i+1}))\right|
		\right\}.
		\label{eq:bp_jacobian}	
	\end{align}
\end{widetext}

We can obtain the correct probability using Eq.~\eqref{eq:bp_naive_correct} instead of Eq.~\eqref{eq:bp_naive}.
The Jacobian $J_{g^{-1}}$ calculation is, however, computationally expensive.

\begin{figure}[tbp]
	\centering
	\includegraphics[width=70mm]{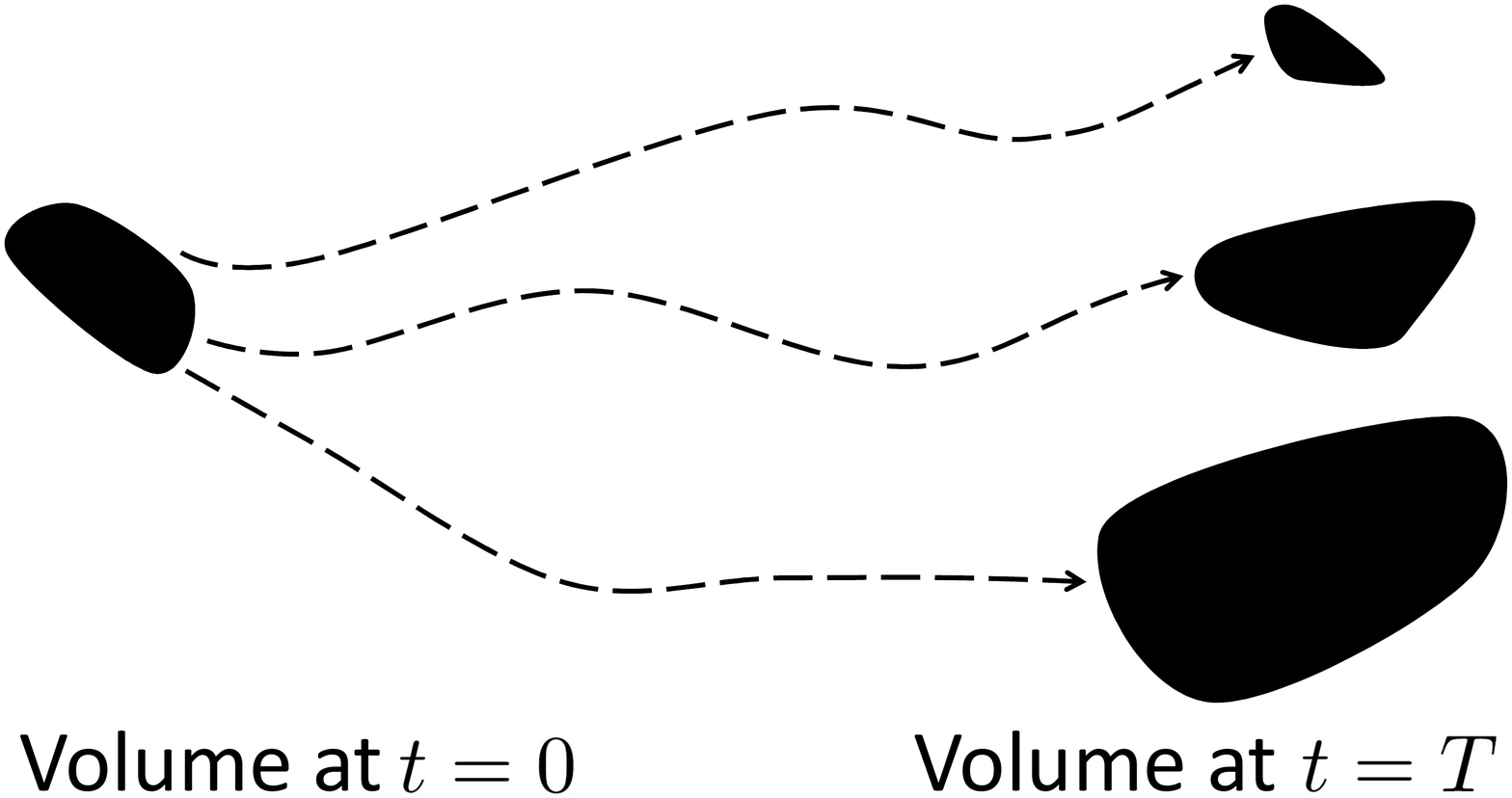}
	\caption{Change in the infinitesimal volume in the state space along each path.}
	\label{fig.factor_limit}
\end{figure}

We note that the factor $J(y_1, \dots, y_N)$ goes to
\begin{equation}
\exp
\left(
-
\int_{0}^{T}
\mathrm{div}{f(x_t)} dt
\right)
\label{eq.factor_limit}
\end{equation}
in the limit as $\Delta t \rightarrow 0$ when we assume that $g(x) = x + f(x)\Delta t$.
The proof of Eq.~\eqref{eq.factor_limit} is given in Appendix \ref{sec:appendix_a}.
This shows that we must include factor  $J(y_1, \dots, y_N)$ for unbiased estimation even in the limit of infinitesimal $\Delta t$.
We can regard the factor written in Eq.~\eqref{eq.factor_limit} as the change in  the infinitesimal volume along each path (Fig.~\ref{fig.factor_limit}).

\section{Time Reverse Monte Carlo Method} \label{sec:trmc}
To overcome these difficulties, we propose the TRMC method.
TMRC essentially involves introducing simplified backward dynamics with a weight.
This weight enables the bias of estimators to be corrected.
First, we introduce a backward transition probability $q\left( x_{i+1} \rightarrow x_{i} \right)$ from $x_{i+1}$ to $x_{i}$.
We can choose an arbitrary probability density $q$, 
while the computation efficiency strongly depends on it.
Once we introduce $q\left( x_{i+1} \rightarrow x_{i} \right)$, 
we can rewrite Eq.~\eqref{eq:fp} as
\begin{widetext}
	\begin{align}
		P(X_N \in A) 
		=
		\int dx_{0:N} 
		\frac{\boldsymbol{1}_{x_N \in A}}{V_A}
		\left\{
		\prod_{i=0}^{N-1}  
		q\left( x_{i+1} \rightarrow x_{i} \right)
		W_{i}
		\right\}
		V_A
		p(x_{0})
		,
		\label{eq:trmc_probability} 
	\end{align}
\end{widetext}
where 
\begin{align}
	W_{i} 
	=
	\frac{\
		p\left( x_{i+1} | x_{i} \right)
	}{
		q\left( x_{i+1} \rightarrow x_{i} \right)
	}
	\label{eq:weight}
\end{align}
is the weight required to correct the bias of estimators 
and $V_A$ is the volume of target region $A$.
Suppose $p(x_0)$ is uniformly distributed on $B \subset \mathbb{R}^D$;
$p(x_0)=\frac{1}{V_B}\boldsymbol{1}_{x_0 \in B}$, $V_B$ is the volume of $B$.
The efficiency of our algorithm does not depend on the factor $V_A$ when $V_B$ is considerably large.
This is the advantage of using our algorithm.

The algorithm consists of the following steps.

\noindent{\bf TRMC Algorithm}
\begin{enumerate}
	\renewcommand{\labelenumi}{Step \arabic{enumi}: }
	\item Draw $M$ samples $\left\{ x_{N}^{(1)}, \cdots, x_{N}^{(M)}\right\}$ from the uniform distribution in $V_A$.
	\item Apply the following steps for $j=1, \dots, M$, and for $i = N-1, \dots, 0$.
	\begin{enumerate}
		\item Generate a sample from $x^{(j)}_{i+1}$ to $x^{(j)}_{i}$ with transition probability 
		$q\left( x^{(j)}_{i+1} \rightarrow x^{(j)}_{i} \right)$.
		\item Calculate weight $W_{i}^{(j)}$ using Eq.~\eqref{eq:weight}.
	\end{enumerate}
	\item Evaluate the unbiased estimates of probability $P(X_N \in A)$ as 
	\begin{equation}
	P(X_N \in A)
	\backsimeq 
	\frac{1}{M} \sum_{j=1}^{M} 
	W^{(j)}	
	,
	\label{eq:trmc_mc}
	\end{equation}
	where the factor
	\begin{equation}
	W^{(j)}
	=
	\left\{
	\prod_{i=0}^{N-1}  
	W_{i}^{(j)}
	\right\}
	V_A
	p(x_{0}^{(j)})
	\label{eq:trmc_factor}
	\end{equation}
	is attached to each simulation path.
\end{enumerate}
The inputs of our algorithm are the number of Monte Carlo paths $M$, 
the number of time steps $N$, 	
the initial distribution $p(x_0)$,
the target region $A$,
and the transition probability density $q$.
When we actually perform the simulation on our computers, we take the logarithm of these weights to prevent numerical overflow.

This algorithm provides unbiased estimates of the desired probabilities.
The idea of this scheme is a kind of SIS~\cite{doucet2001introduction}.
An advantage of our method is that we do not need to calculate $g^{-1}$ or their Jacobian matrices at each $i$.

The remaining problem involves determining the method for choosing the transition probability $q\left( x_{i+1} \rightarrow x_{i} \right)$. 
The basic idea is to choose the backward dynamics that generates trajectories similar to the forward dynamics defined by Eq.~\eqref{eq:forward_general}.
The similarity of the trajectory is measured by $W^{(j)}$ in Eq.~\eqref{eq:trmc_factor}.

\section{Implementation for Stochastic Difference Equation} \label{sec:isde}
To give concrete examples of the transition probability 
$q\left( x_{i+1} \rightarrow x_{i} \right)$, 
we assume the forward dynamics to be given in the following form:
\begin{align}
	X_{i+1} = X_{i} + f\left(X_{i} \right)\Delta t + \epsilon_{i} \sqrt{\Delta t}.
	\label{eq:fd}
\end{align}
This corresponds to the case wherein $g(x) = x + f(x)\Delta t$ in Eq.~\eqref{eq:forward_general}. 
The noise $\epsilon_i$ is assumed to be i.i.d. Gaussian noise with mean zero and the variance-covariance matrix $\Sigma = \sigma \sigma^T$.
This class of equations appears in a wide range of problems in many different fields such as physics~\cite{heermann1990computer}, computational chemistry~\cite{vanden2010transition}, and mathematical finance~\cite{joshi2003concepts, oksendal2013stochastic}.

In this case, as a simple choice, we can use the following backward dynamics:
\begin{align}
	X_{i} = X_{i+1} - f\left(X_{i+1} \right)\Delta t + \epsilon_{i} \sqrt{\Delta t}.
	\label{eq:bd}
\end{align}
This approximation corresponds to 
substituting $f\left(X_{i+1} \right)$ for $f\left(X_{i} \right)$ in Eq.~\eqref{eq:fd}.

With this choice, weight $W_i$ in Eq.~\eqref{eq:weight} takes the form
\begin{widetext}
	\begin{align}
		W_{i}
		&=
		\frac{
			p\left( x_{i+1} | x_{i} \right)
		}{
			q\left( x_{i+1} \rightarrow x_{i} \right)
		}
		=
		\frac{
			\exp
			\left[
			-\frac{1}{2}
			\left(x_{i+1} - x_{i} -  f(x_{i})\Delta t \right)^{T}
			\left(\Sigma \Delta t\right)^{-1}
			\left(x_{i+1} - x_{i} -  f(x_{i})\Delta t \right)
			\right]
		}{
			\exp
			\left[
			-\frac{1}{2}
			\left(x_{i+1} - x_{i} -  f(x_{i+1})\Delta t \right)^{T}
			\left(\Sigma \Delta t\right)^{-1}
			\left(x_{i+1} - x_{i} -  f(x_{i+1})\Delta t \right)
			\right]
		} \nonumber \\
		&=
		\exp
		\left[
		-\left( f(x_{i+1}) - f(x_{i}) \right)^{T}
		\Sigma^{-1}
		\left(
		(x_{i+1} - x_{i})
		-
		\frac{\Delta t}{2}
		\left(
		f(x_{i+1}) + f(x_{i})
		\right)
		\right)
		\right]
		.
		\label{eq.ww_iba}
	\end{align}
\end{widetext}
As we show in the next section, the resultant algorithm is simple yet effective compared with the forward simulation when the target region $A$ is smaller than the support of the initial distribution $p(x_0)$.

We note that the factor $\prod_{i=0}^{N-1} W_{i}$ goes to
\begin{equation}
\exp
\left(
-
\int_{0}^{T}
\mathrm{div}{f(x_t)} dt
\right)
\label{eq.factor_limit_2}
\end{equation}
in the limit as $\Delta t \rightarrow 0$.
The proof of Eq.~\eqref{eq.factor_limit_2} is given in Appendix \ref{sec:appendix_b}.
Note that Eq.~\eqref{eq.factor_limit_2} coincides with Eq.~\eqref{eq.factor_limit} derived from a different assumption.


\section{Experimental Results}
\label{sec:er}
We present the numerical results in this section.
Forward simulations (FS) are used to check the consistency and computational efficiency of our result.

Using forward and backward dynamics, we simulate sample trajectories 
$x=\left\{x_1, \cdots, x_N \right\}$ generated by each model and 
compute the probability $P(X_N \in A)$ from $M$ independent simulations.

We denote a standard error of TRMC to evaluate the computational efficiency by $\sigma_{s}$.
We also denote the standard error of FS by $\sigma_{s}^{F}$.
Using these variables, we define a relative value of variance by 
\begin{align}
	\rho_{1}
	=
	\left(
	\frac{\sigma_{s}^{F}}{\sigma_s}
	\right)^2
	.
\end{align}
The factor $\rho_{1}$ indicates the computational efficiency only including the effect caused by the variance of estimators for a fixed sample size.
With this definition, more complex algorithms tend to be more efficient while they require more computational time.
Then, we also define another measure of the relative computational efficiency $\rho_2$ as
\begin{align}
	\rho_{2}
	=
	\rho_{1}
	\frac{\tau^{F}}{\tau},
\end{align}
where $\tau$ is the computational time in seconds of the simulation and $\tau^{F}$ is the computational time of FS in seconds. 
This efficiency is defined in the sense of the actual performance considering both the computational time and the variance of the resulting estimates.

\subsection{Stochastic typhoon model}
\label{subsec:stochastic_typhoon}
The first example is a stochastic typhoon model~\cite{Nakano2013}, which gives an example of risk estimation by the proposed method. 
The stochastic typhoon model was designed to reproduce the statistics of typhoons in the northwestern
part of the Pacific Ocean.
This is a four-dimensional model given by
\begin{align}
	\begin{aligned}
		x_{i+1} &= x_{i} + v_{i}, \\
		v_{i+1} &= V\left(x_{i+1} \right)  +
		w \left( v_{i} - V\left(x_{i} \right) \right)
		+ 
		\epsilon_i
		,\\ 
		V\left(x_{i} \right)  &= a_{0} + a_{1}x_{\phi, i} + a_{2}\sin x_{\lambda, i} + a_{3}\sin^2 x_{\lambda, i},  
	\end{aligned}
	\label{eq:sde_typhoon}
\end{align}
where we use a global coordinate system defined by the geographic longitude ($\phi$) and latitude ($\lambda$).
We also define the two-dimensional position $x = \left(x_{\phi}, x_{\lambda}\right)$, speed $v = \left(v_{\phi}, v_{\lambda}\right)$ of a typhoon, and function $V(x) = \left(V_{\phi}(x), V_{\lambda}(x) \right)$. 
$w, a_0, a_1, a_2$, and $a_3$ are constants.
The noise $\epsilon$ obeys a Gaussian distribution with mean zero and variances $\sigma^2$.

We fix $w=0.93$, $a_{0}=(0.792, 0.538)$, $a_{1}=(0.122, 0.371)$, $a_{2}=(-0.513, 0.583)$, $a_{3}=(0.770, -0.387)$, $\sigma=0.4$.
The target region $A$ is $\left\{(x_{\phi}, x_{\lambda}); 138.5 \le x_{\phi} \le 139.5, 34.5 \le x_{\lambda} \le 35.5 \right\}$.
Since there is no range constraint on the distribution of the final speed $v_f$ at the target, 
we adopt a uniform distribution with a suitably wide range $U_f$; here, $U_f$ is defined as the region 
$
\bigl\{
(v_{\phi}, v_{\lambda}); 
V_{\phi}(x_A)    -3 \le v_{\phi}    \le V_{\phi}(x_A)+3,  
V_{\lambda}(x_A) -3 \le v_{\lambda} \le V_{\lambda}(x_A)+3 
\bigr\},
$
where $x_A$ is the center of target region A.

\begin{figure*}[ht]
	\begin{tabular}{cc}
		\begin{minipage}[t]{0.45\hsize}
			\centering
			\includegraphics[width=70mm, height=60mm]{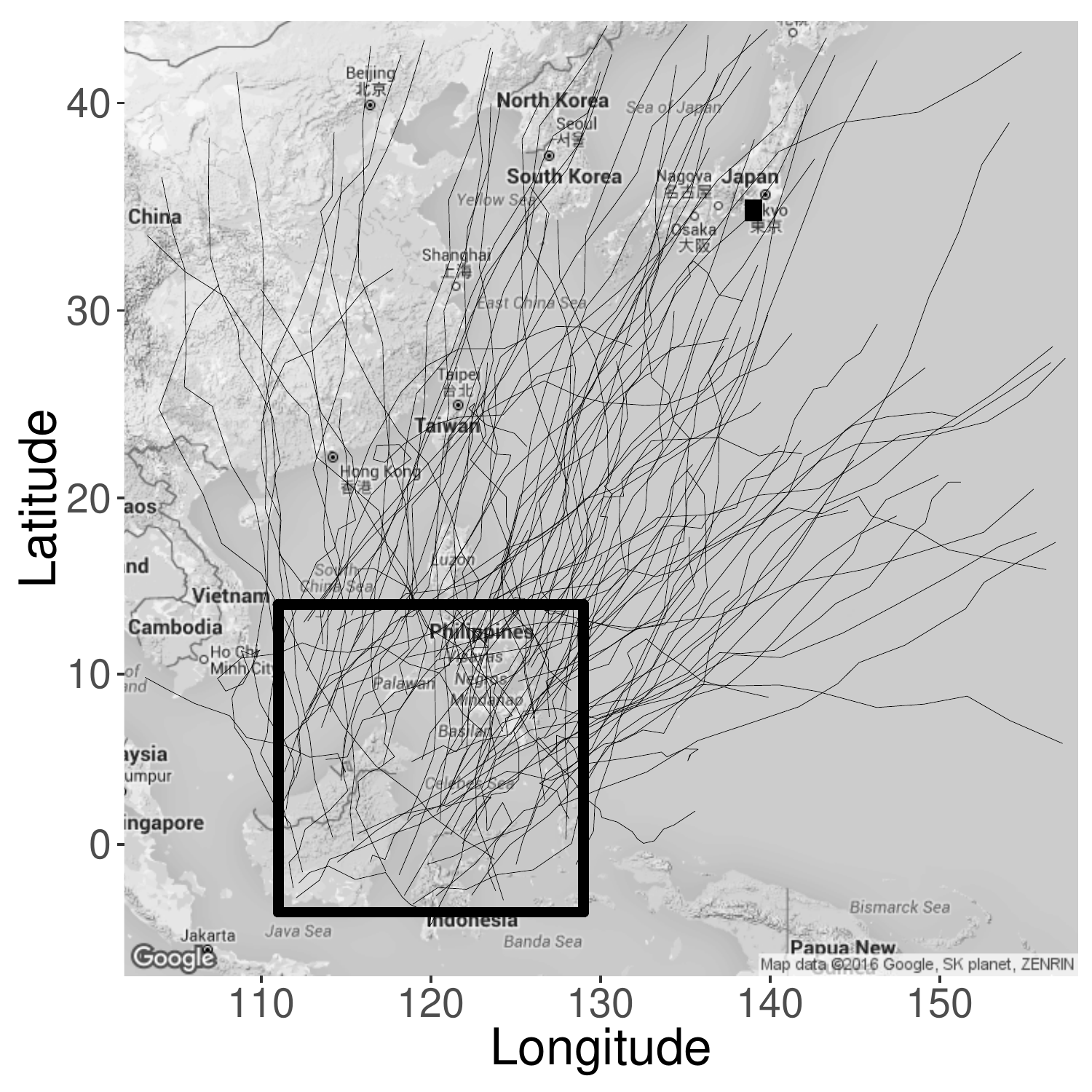}
			\caption{Example of Monte Carlo paths generated by the stochastic typhoon model originating from the northwestern part of the Pacific Ocean. Each line corresponds to a path generated by the forward simulation. The black rectangular region shows the possible initial position of typhoons in the northwestern part of the Pacific Ocean. The initial positions of typhoons are uniformly distributed.}
			\label{fig.typhoon_f}
		\end{minipage} 
		&
		\begin{minipage}[t]{0.45\hsize}
			\centering
			\includegraphics[width=70mm, height=60mm]{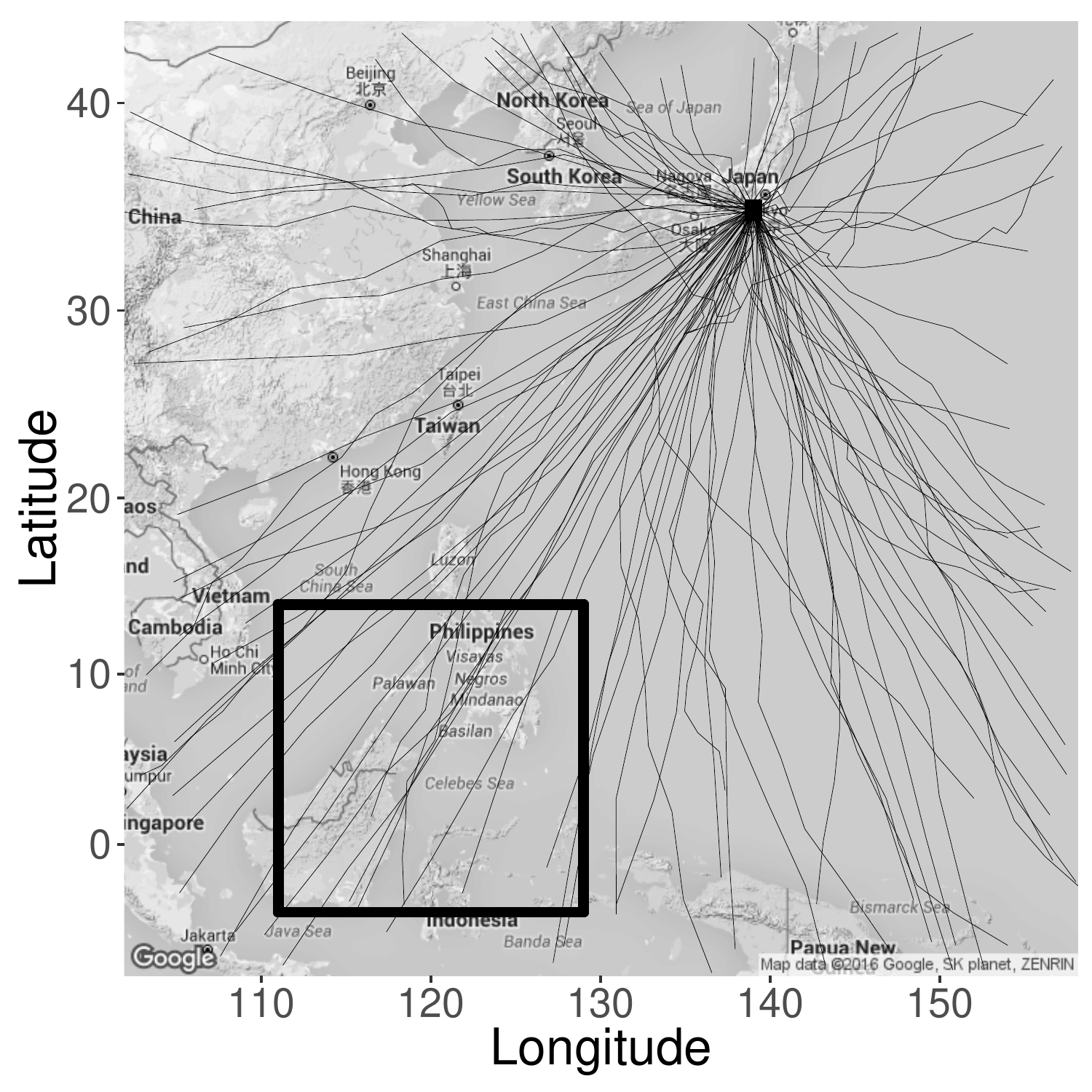}
			\caption{Example of Monte Carlo paths generated by TRMC starting from Tokyo. 
				Each line corresponds to a path generated by TRMC.
				The black rectangular region corresponds to the possible initial 
				position of typhoons in the northwestern part of the Pacific Ocean.
			}
			\label{fig.typhoon_b}
		\end{minipage} 
	\end{tabular}
\end{figure*}

We also assume that the initial condition is uniformly distributed in
$
D
=
\bigl\{
(x, v); 
111 \le x_{\phi} \le 129, 
-4 \le x_{\lambda} \le 14, 
v \in U_0
\bigr\},
$
where $U_0$ is defined as the region 
$
\bigl\{
(v_{\phi}, v_{\lambda}); 
V_{\phi}(x_0)    -1.5 \le v_{\phi}    \le V_{\phi}(x_0) + 1.5,  
V_{\lambda}(x_0) -1.5 \le v_{\lambda} \le V_{\lambda}(x_0) + 1.5 
\bigr\}
$
and 
$x_0 = (120, 5).$
This corresponds to the case wherein typhoons that occurred in the Philippines travel to the Tokyo area exactly with a small probability. 
We set $M$ to $10^{8}$ and $N$ to $16$.
Examples of Monte Carlo paths for both simulations are given in Figs.~\ref{fig.typhoon_f} and \ref{fig.typhoon_b}.

This simulation was carried out on a laptop computer with 2.3 Ghz Intel core i5 and 8 GBytes memory. 
The computational time of TRMC in this simulation for generating $10^8$ Monte Carlo paths is around $4.0 \times 10^3$ seconds.

\begin{figure}[tbp]
	\centering
	\includegraphics[width=80mm]{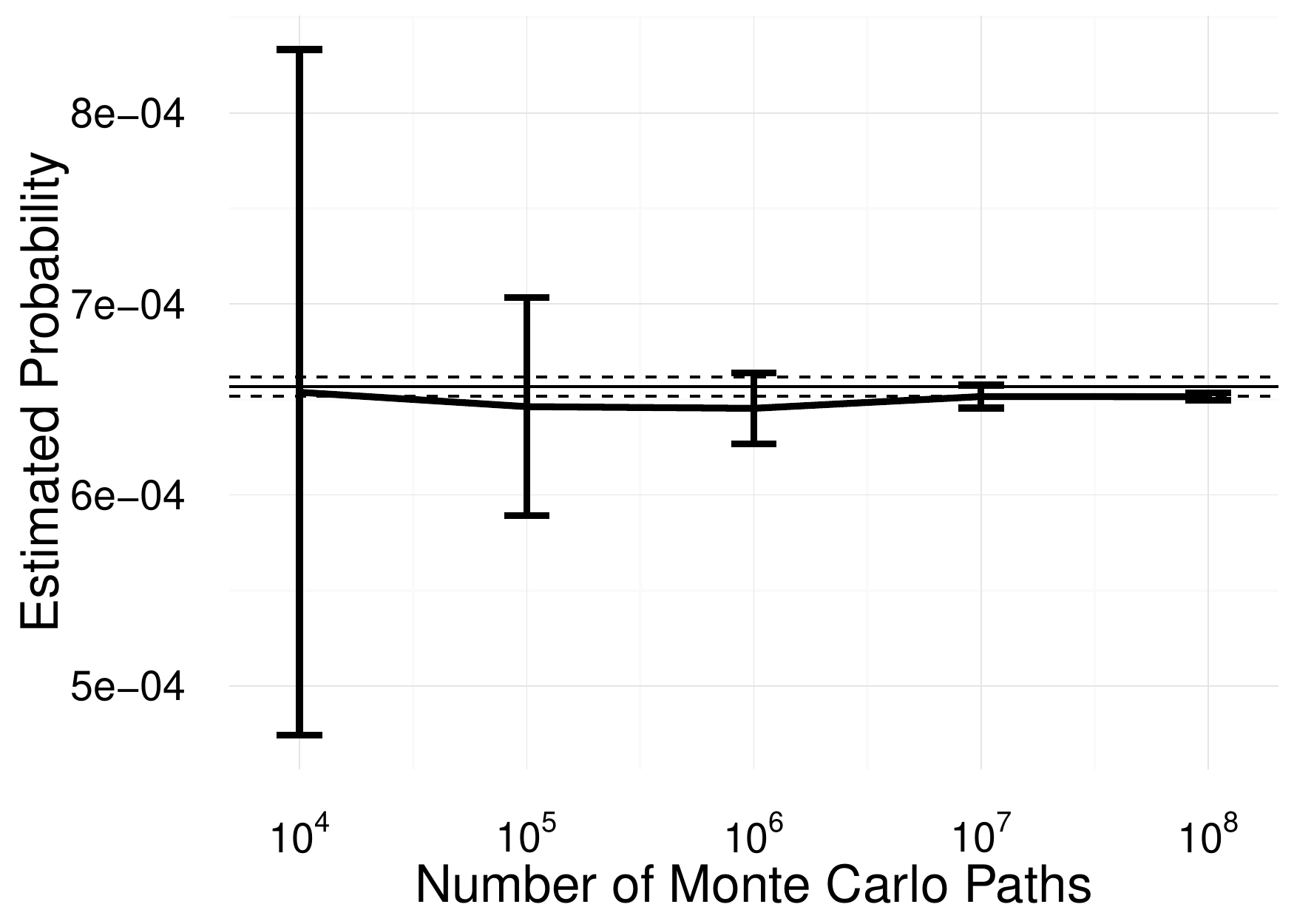}
	\caption{Convergence of TRMC for the stochastic typhoon model.
		The estimated probabilities converge to those obtained by FS as the number of Monte Carlo paths increases. 
		Error bars indicate approximate $\pm 1$ standard error confidence intervals for TRMC.
		The horizontal solid line indicates the estimated probability by FS. 
		The horizontal dashed line represents $\pm 1$ standard error confidence intervals for FS.
		FS has the same number of Monte Carlo paths, $M=10^8$, as TRMC.		
	}
	\label{fig.typhoon_convergence_1}
\end{figure}

Table \ref{tab:typhoon_1} shows the result of computational experiments for the stochastic typhoon model.
It shows that the probabilities of FS and TRMC agree within the error bars.
If we ignore the factor defined by Eq.~\eqref{eq:trmc_factor}, it does not reproduce the unbiased probability as in the case of the stochastic difference equation.
Furthermore, it shows that TRMC is 4.2 times in terms of $\rho_2$ and 7.3 times in $\rho1$) more efficient than FS. 
Fig.~\ref{fig.typhoon_convergence_1} shows the convergence of TRMC when the number of Monte Carlo paths $M$ increases. 
It reveals that our algorithm converges correctly on increasing the number of Monte Carlo paths $M$.

To simulate events with smaller probabilities, we make the target region $A$ smaller as 
$\left\{(x_{\phi}, x_{\lambda}); 138.75 \le x_{\phi} \le 139.25, 34.75 \le x_{\lambda} \le 35.25 \right\}$.
It shows that the smaller the probability, the more efficient our algorithm becomes as compared with the FS.

\begin{figure}[htb]
	\centering
	\includegraphics[width=70mm, height=60mm]{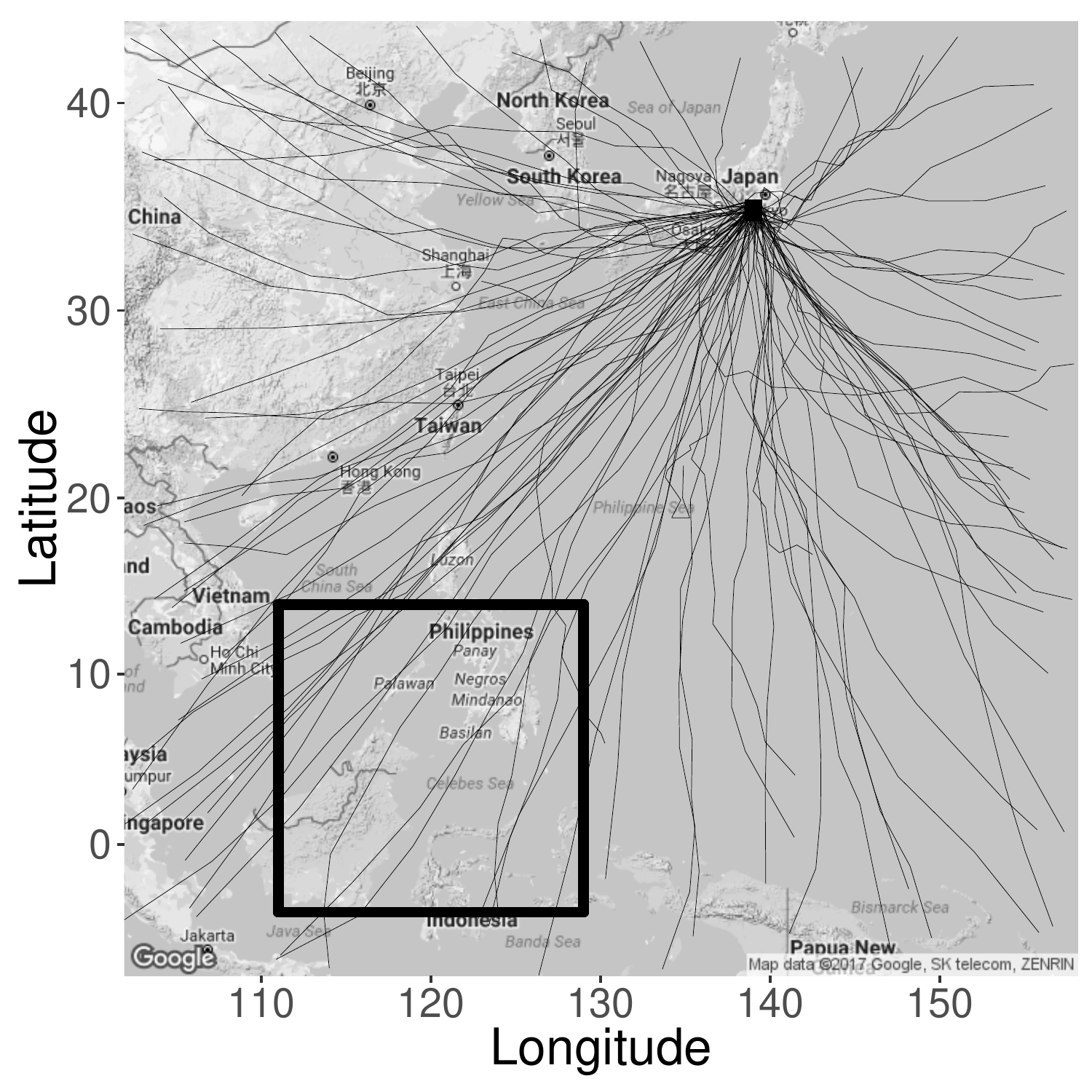}
	\caption{Example of Monte Carlo paths generated by TRMC starting from Tokyo.
		The velocity distribution is restricted to the tendency to move southward.
		Each line corresponds to a path generated by TRMC.
		The black rectangular region corresponds to the possible initial 
		position of typhoons in the northwestern part of the Pacific Ocean.
	}
	\label{fig.typhoon_b2}
\end{figure}

\begin{figure}[tbp]
	\centering
	\includegraphics[width=80mm]{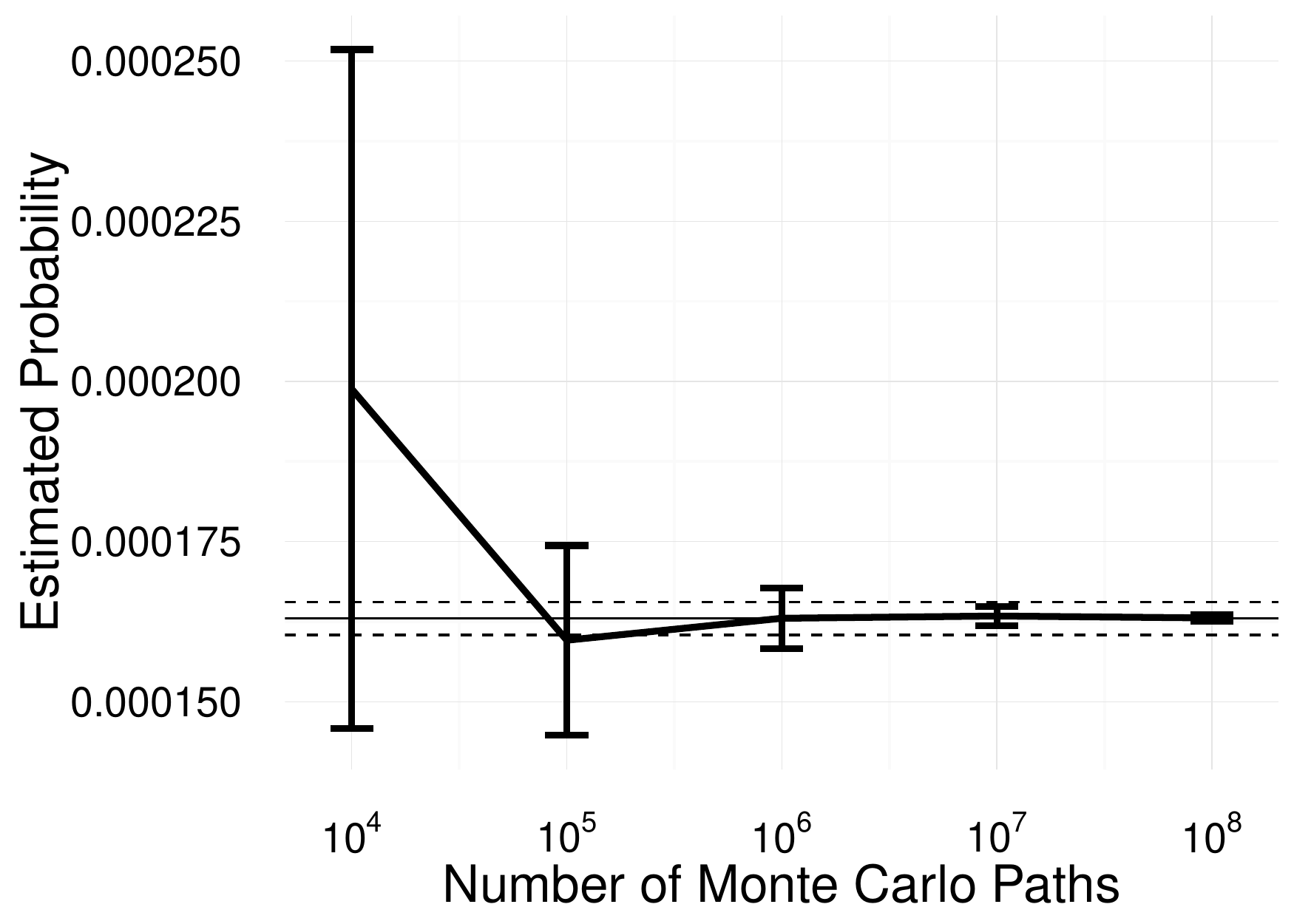}
	\caption{
		Convergence of TRMC for the stochastic typhoon model in the
		smaller probability case.
		The estimated probabilities converge to those obtained by FS as the number of Monte Carlo paths increases. 
		Error bars indicate approximate $\pm 1$ standard error confidence intervals for TRMC.
		The horizontal solid line indicates the estimated probability by FS. 
		The horizontal dashed line represents $\pm 1$ standard error confidence intervals for FS.
		FS has the same number of Monte Carlo paths, $M=10^8$, as TRMC.		
	}
	\label{fig.typhoon_convergence_2}
\end{figure}

In Fig.~\ref{fig.typhoon_b}, a few Monte Carlo paths are shown to have moved northward.
To prevent this from happening and improve its efficiency, we restrict the velocity distribution of Monte Carlo paths to the tendency to move southward.
We change the range $U_f$ of the final speed $v_f$ to 
$
\bigl\{
(v_{\phi}, v_{\lambda}); 
V_{\phi}(x_A)    -3 \le v_{\phi}    \le V_{\phi}(x_A)+3,  
V_{\lambda}(x_A) -2 \le v_{\lambda} \le V_{\lambda}(x_A)+2 
\bigr\}.
$
We call this simulation TRMC (restricted) in Fig.~\ref{fig.typhoon_b2}.
Table \ref{tab:typhoon_1} shows that the probabilities of TRMC and TRMC (restricted) agree within error bars.
Because the number of unnecessary Monte Carlo paths moving northward decreases, TRMC (restricted) is more efficient than TRMC. 
More severe constraint $v_{\lambda} \ge 0$, however, causes a small bias in the estimated probabilities: 
see TRMC ($v_{\lambda} \ge 0$) in Table \ref{tab:typhoon_1}.
Fig.~\ref{fig.typhoon_convergence_2} also shows that our algorithm converges correctly on increasing the number of Monte Carlo paths $M$.

So far, we consider the case that the number of time steps from the initial position to Tokyo ($N=16$) is precisely known. 
We can relax this assumption, but we should be careful with a limitation of a discrete time model.
A typhoon can pass nearby Tokyo, for example, between $N=15$ and $N=16$, which causes an underestimation of the actual risk when we only consider hit at integral time steps. 
A way to reduce this effect is to develop models with smaller steps, while it is also possible to
introduce some initialization or interpolation method into a backward simulation.
However, we leave this as a future problem, because this study aims to check whether the concept of backward simulation is mathematically valid.

\begin{table}[htb]
	\centering
	\caption{Comparison among TRMC, TRMC (restricted), TRMC (no weight), and FS for stochastic typhoon model.}
	\begin{tabular}{ccccc}
		\multicolumn{4}{c}{Case I} \\ 
		Method & $P(X_N \in A)$ & $\sigma_{s}$ & $\rho_1$ & $\rho_2$ \\ \hline 
		TRMC  & $6.514 \times 10^{-4}$ & $0.009 \times 10^{-4}$ & $7.3$  & $4.2$ \\ 
		TRMC (restricted) & $6.501 \times 10^{-4}$ & $0.007 \times 10^{-4}$ & $13.5$ & $7.9$ \\ 
		TRMC ($v_{\lambda} \ge 0$)  & $6.424 \times 10^{-4}$ & $0.007 \times 10^{-4}$ & $ - $  & $  -$ \\
		TRMC (no weight)  & $0.805 \times 10^{-4}$ & $0.001 \times 10^{-4}$ & $ - $  & $  -$ \\
		FS                & $6.568 \times 10^{-4}$ & $0.026 \times 10^{-4}$ & $1.0$  & $1.0$ \\ \hline 
	\end{tabular} 
	\label{tab:typhoon_1}
	
	\begin{tabular}{ccccc}
		\multicolumn{4}{c}{Case II} \\ 
		Method & $P(X_N \in A)$ & $\sigma_{s}$ & $\rho_1$ & $\rho_2$ \\ \hline
		TRMC            & $1.631 \times 10^{-4}$ & $0.002 \times 10^{-4}$ & $29.0$ & $16.4$ \\
		TRMC (no weight) & $0.202 \times 10^{-4}$ & $0.000 \times 10^{-4}$ & $  -$  & $   -$ \\ 
		FS              & $1.630 \times 10^{-4}$ & $0.012 \times 10^{-4}$ & $1.0$  & $ 1.0$ \\ \hline 
	\end{tabular}
	\label{tab:typhoon2}
\end{table}

\subsection{Lorenz 96 Model}
\label{subsec:lorenz}
As a higher-dimensional example, we evaluate the efficiency of our algorithm for the Lorenz 96 model~\cite{lorenz1996predictability, wouters2015rare}.
The Lorenz 96 model is an atmospheric model and was introduced by Edward Lorenz in 1996.
It is defined as the set of coupled ordinary differential equations

\begin{align}
	\frac{dx_k}{dt} &= f_{k}(x) + \epsilon_{k}, \nonumber  \\
	f_{k}(x) &= -x_{k-2}x_{k-1} + x_{k-1}x_{k+1} - x_{k} + F,  \label{eq:lorenz96_equation} \\
	k &= 1\dots K, \nonumber 
\end{align}
where $x=\left\{x_k; k=1\dots K \right\}$ is the state of the system and $F$ is a constant. 
We set $K=9$ and introduce Gaussian noise $\epsilon_{k}$ with mean zero and variance $\sigma^2$.
Here, we choose $F=8$, a value known to cause weak chaotic behavior and often used as a benchmark in data assimilation~\cite{ott2004local}. 

To simulate Eq.~\eqref{eq:lorenz96_equation}, 
we have to discretize it.
While many discretization schemes are available, we focus on the simplest and most common scheme, the Euler scheme.
The time-discretized version of Eq.~\eqref{eq:lorenz96_equation} by the Euler scheme is 

\begin{equation}
x_{k, i+1} = x_{k, i} + f(x_i)\Delta t + \epsilon_{k} \Delta t, k=1 \dots K,
\label{eq:lorenz96_equation_discretized}
\end{equation}
where we set $\Delta t$ to $0.001$ and $\sigma$ to $0.1/\sqrt{\Delta t}$.  

The target region $A$ is 
$ \left\{(x_1, \dots, x_K)| -5.0 \le x_i \le 7.0; i=1 \dots K \right\}$. 
We also assume that the initial state for ${x_i}$ is uniformly distributed in $D=\left\{(x_1, \dots, x_K)| 1.5 \le x_i \le 8.5; i=1 \dots K \right\}$. 
We set $M$ to $10^7$ and $N$ to $100$. 

We conduct this simulation on the environment described in Sect.~\ref{subsec:stochastic_typhoon}.
The computational time of TRMC in this simulation for generating $10^7$ Monte Carlo paths is around $3.0 \times 10^3$ seconds.

\begin{figure}[tbp]
	\centering
	\includegraphics[width=80mm]{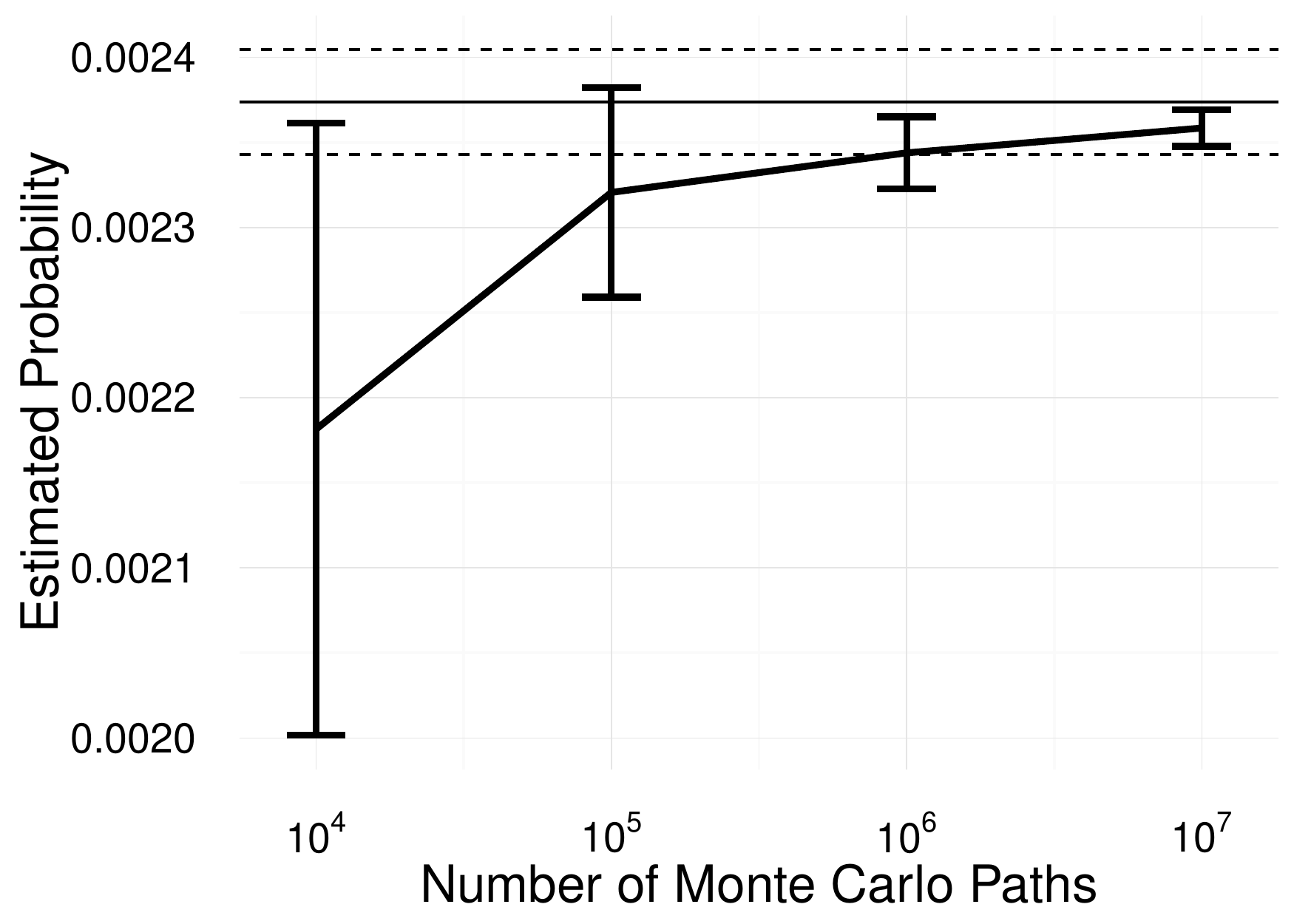}
	\caption{Convergence of TRMC for the Lorenz 96 model.
		The estimated probabilities converge to those obtained by FS as the number of Monte Carlo paths increases. 
		Error bars indicate approximate $\pm 1$ standard error confidence intervals for TRMC.
		The horizontal solid line indicates the estimated probability by FS. 
		The horizontal dashed line represents $\pm 1$ standard error confidence intervals for FS.
		FS has the same number of Monte Carlo paths, $M=10^7$, as TRMC.		
	}
	\label{fig.lorenz96_convergence}
\end{figure}

Table \ref{tab:lorenz96-1} shows the result of computational experiments for the Lorenz 96 model. 
It shows that the probabilities of TRMC and FS agree within the error bars.
The case where we ignore the factor defined by Eq.~\eqref{eq:trmc_factor} does not reproduce the same unbiased probability as the other computational experiments.
The result shows that TRMC can perform better for estimating the probabilities in the  high-dimensional case.
TRMC is 5.18 times more efficient than FS in terms of $\rho_2$, and 8.23 times in $\rho1$ more efficient in Table \ref{tab:lorenz96-1}. 
Fig.~\ref{fig.lorenz96_convergence} shows the convergence of TRMC when the number of Monte Carlo paths $M$ increases. 
It reveals that our algorithm converges correctly on increasing the number of Monte Carlo paths $M$.

\begin{table}[htb]
	\centering
	\caption{Comparison among TRMC, TRMC (no weight), and FS for the Lorenz 96 model.}
	\begin{tabular}{ccccc}
		Method & $P(X_N \in A)$ & $\sigma_{s}$  & $\rho_1$ & $\rho_2$ \\ \hline
		TRMC            & $2.358 \times 10^{-3}$ & $0.005 \times 10^{-3}$ & $8.23$ & $5.18$ \\
		TRMC (no weight) & $0.957 \times 10^{-3}$ & $0.001 \times 10^{-3}$ & $   -$ & $   -$ \\
		FS              & $2.373 \times 10^{-3}$ & $0.015 \times 10^{-3}$ & $1.00$ & $1.00$ \\ \hline 
	\end{tabular} 
	\label{tab:lorenz96-1}
\end{table}

\section{Improved Scheme with Resampling}
\label{sec:is}\label{sec:is_resampling}
\begin{figure*}[tbp]
	\begin{tabular}{c}
		\begin{minipage}[t]{0.45\hsize}
			\centering
			\includegraphics[width=80mm]{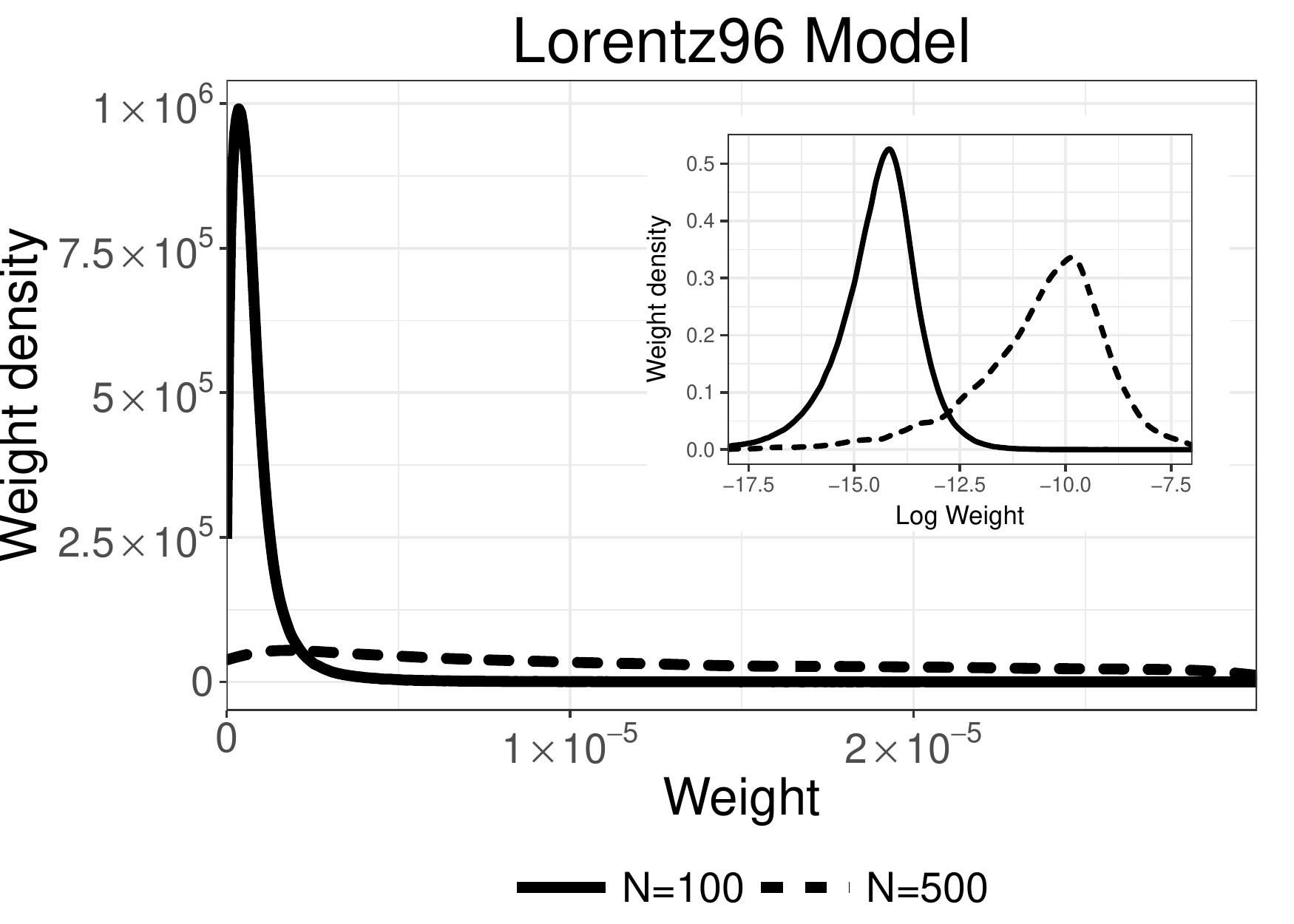}
		\end{minipage}
	\end{tabular}
	\caption{
		Distribution of weight $\prod_{i=0}^{N-1} W_i$ in TRMC with different numbers of time steps.
		The vertical and horizontal lines indicate the weight density
		and the value of weights, respectively.
		The weight distributions with a large number of time steps have a heavy-tailed distribution.}
	\label{fig.weight_comparison}
\end{figure*}

\begin{figure}[tbp]
	\centering
	\includegraphics[width=80mm]{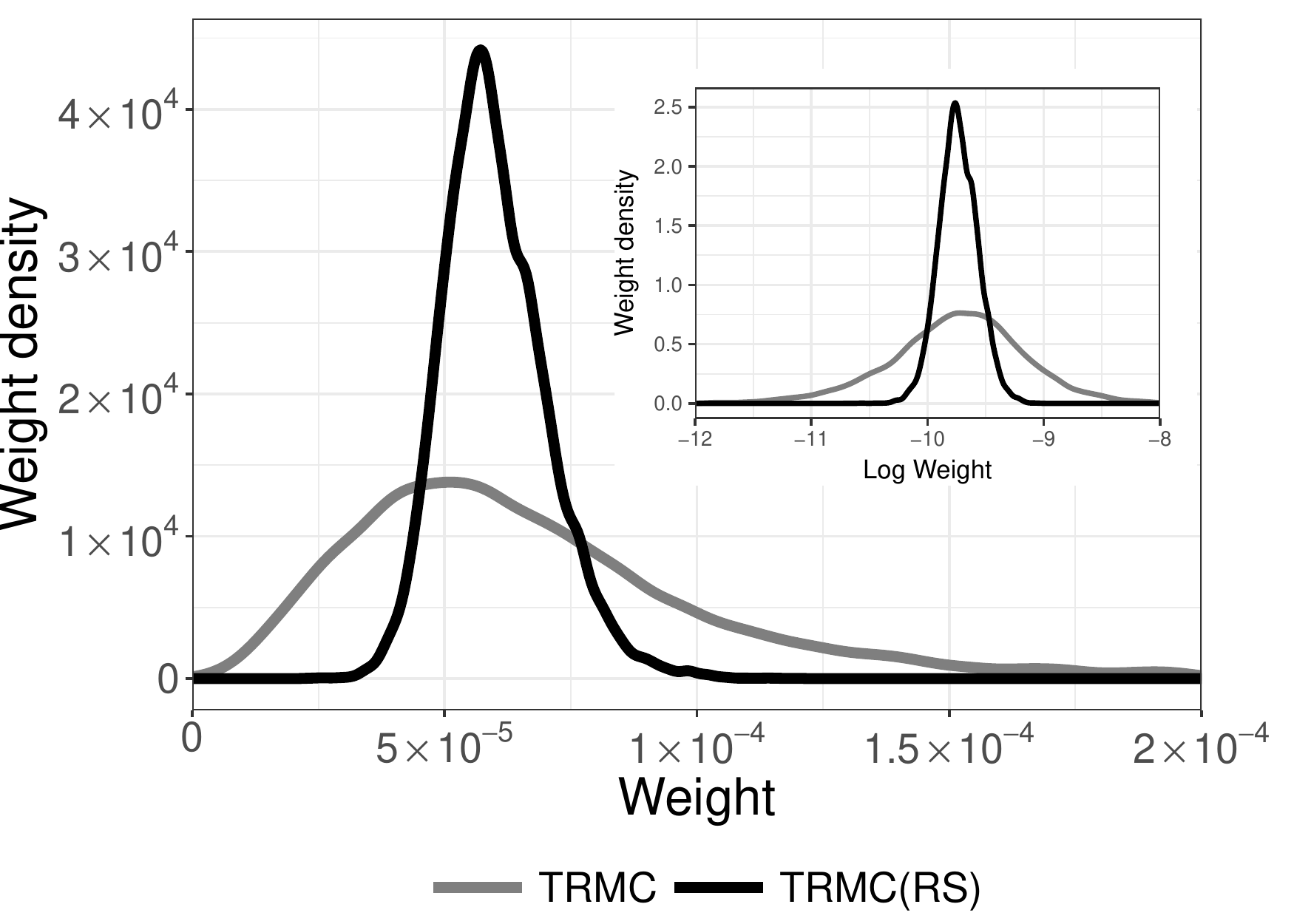}
	\caption{Distribution of the weight $\prod_{i=0}^{N-1} W_i$ in TRMC and TRMC (RS) for the Lorenz 96 model.	
		We set the threshold $\alpha$ to $0.5$ for TRMC (RS).
		The variance of the distribution is much smaller for TRMC (RS) than for TRMC.
	}
	\label{fig.resampling_weight_lorenz96}
\end{figure}

Let us consider cases with a larger number of time steps.
The proposed algorithm may not always work efficiently in this situation.
For example, we consider the case where $N$ is equal to $500$ in the Lorenz 96 model (Fig.~\ref{fig.weight_comparison}); these weights are normalized such that their sum is $1$, i.e., $\sum _{j=1}^{M}W^{(j)} = 1$.
The inset located at the top right of the figure shows the graph with a logarithmic scale on the x-axis.
This style is also used in Fig.~\ref{fig.resampling_weight_lorenz96}.
The weight distribution corresponding to  $N=500$  in Fig.~\ref{fig.weight_comparison} has a heavy-tailed distribution.
This phenomenon is referred to as degeneracy, and it means that the weights become unbalanced, and a few weights dominate all the others.
This consequently causes a decrease in computational efficiency~\cite{2001sequential}. 

We introduce an improved scheme to solve this problem, which is realized by resampling~\cite{doucet2009tutorial, tailleur2007probing, zuckerman2010statistical, tuanase2003topological}.
Hereafter, we denote it as TRMC (RS).
This algorithm is effective when both the number of time steps and the amount of noise are large.

Note that our algorithm is based on time-reversed dynamics 
and uses SMC differently from the previous studies 
~\cite{tailleur2007probing, laffargue2013large, wouters2015rare, allen2009forward, zuckerman2011equilibrium} 
on rare event sampling.

We assume that the resampling procedure modifies the weight at $s$ time step
\begin{equation}
\prod_{i=0}^{s-1} W_{i}
\end{equation}
of each Monte Carlo path to an unweighted one by eliminating Monte Carlo paths having small weights and by multiplying
Monte Carlo paths having large weights.

We denote the $j$th Monte Carlo path as $x^{(j)}= \left\{x^{(j)}_{0}, \dots, x^{(j)}_{s} \right\}$.
The procedure of resampling is as follows:
\begin{enumerate}
	\item Define normalized weights 
	$$\tilde{W}^{(j)} = \frac{\prod_{i=0}^{s-1}W_{i}^{(j)}}{\sum_{j=1}^{M} \prod_{i=0}^{s-1} W_{i}^{(j)}}.$$
	\item Resample $M$ times with replacement from set $\left\{ x^{(j)}\right\}_{j=1}^{M}$ of Monte Carlo paths, where the probability of sampling set of $x^{(j)}$ is proportional to $\tilde{W}^{(j)}$.
\end{enumerate}
After a resampling step, 
Monte Carlo paths $\left\{x^{(j)}\right\}_{j=1}^{M}$ 
and associated weights
$\left\{W^{(j)} \right\}_{j=1}^{M}$ 
are replaced by the set of replicated Monte Carlo paths with an equal importance weight
$W^{(j)}=\frac{1}{M}\sum_{j=1}^{M}\prod_{i=0}^{s-1}W_{i}^{(j)}$. 
Degeneracy is estimated using the effective sample size~\cite{kong1994sequential}:
\begin{equation}
M_{eff} = \frac{1}{\sum_{j=1}^{M} (\tilde{W}^{(j)})^2}.
\end{equation}
A small value of $M_{eff}$ corresponds to high degeneracy. 
Hence, a resampling procedure is performed when this value is lower than a certain threshold $\Theta=\alpha M$, where
$\alpha$ is a relative threshold.
That is, a resampling procedure is performed when $\frac{M_{eff}}{M} < \alpha$.

\begin{figure}[tbp]
	\centering
	\includegraphics[width=90mm]{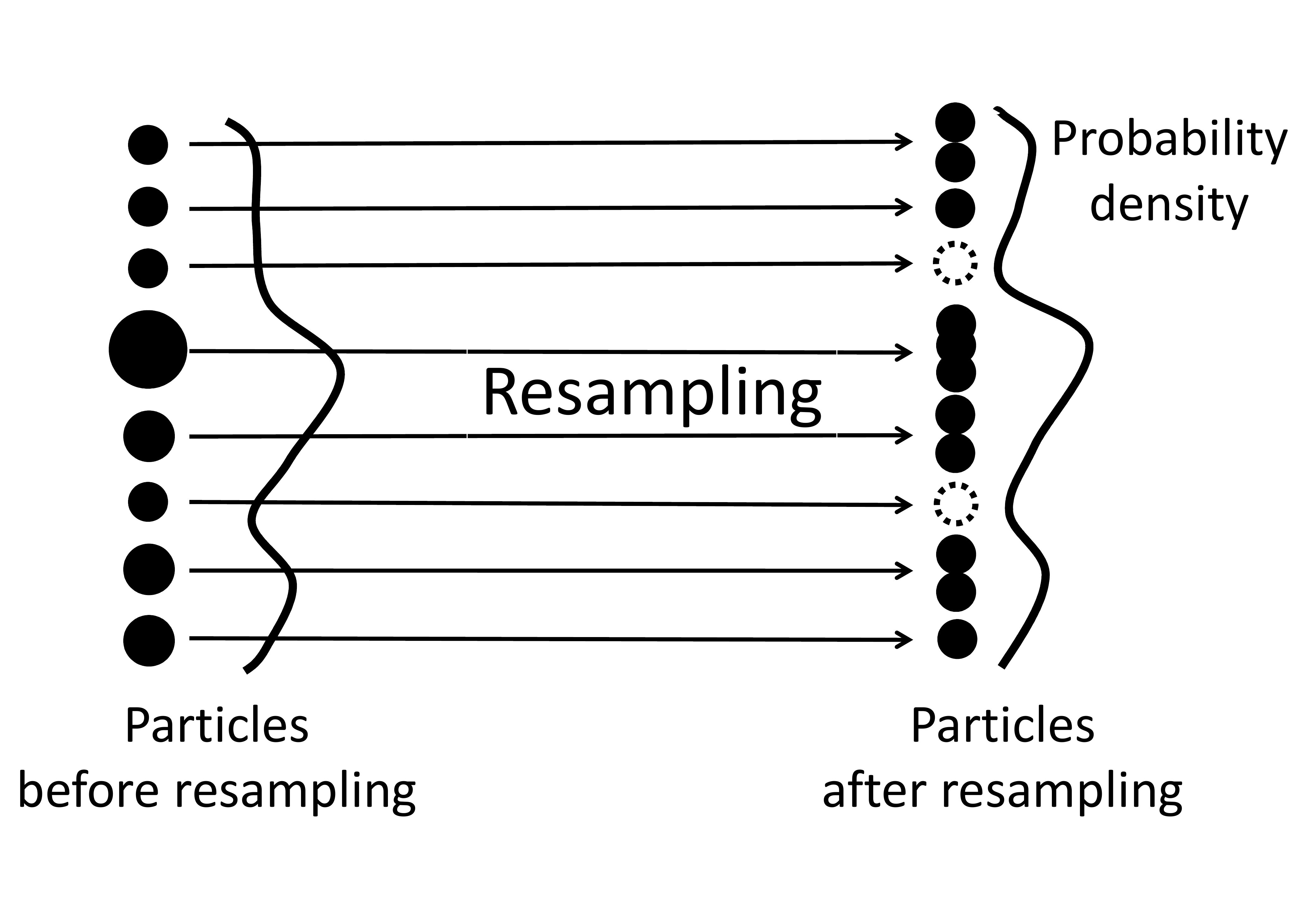}
	\caption{Graphical example of resampling.
		Particles with large weights are replaced with multiple copies of them, 
		and particles with small weights are removed.
	}
	\label{fig:concept_resampling}
\end{figure}

We can use the Eq.~\eqref{eq:trmc_probability} to evaluate probabilities in this case.
Figure~\ref{fig:concept_resampling} shows a graphical scheme of resampling.	

Using this resampling, 
we simulate the Lorenz 96 model with $\sigma=0.3/\sqrt{\Delta t}$, which is larger than that in Sect.~\ref{subsec:lorenz}.
We set the threshold $\alpha$ to $0.05, 0.5$, and $0.9$.
The simulations with these threshold values of $\alpha$ are denoted by $\alpha=5\%,50\%$, and $90$\% respectively.

Table \ref{tab:resampling} shows the result of computational experiments for the Lorenz 96 model.
It shows that the probabilities of FS, TRMC, and TRMC (RS, $\alpha$=50\%) agree within the error bars.

\begin{table}[htb]
	\centering
	\caption{Comparison among TRMC, TRMC (RS, $\alpha$=50\%), and FS for the Lorenz 96 model.}    
	\begin{tabular}{ccccc}
		Method & $P(X_N \in A)$ & $\sigma_{s}$  & $\rho_1$ & $\rho_2$ \\ 
		\hline
		TRMC                     & $2.500 \times 10^{-3}$ & $0.025 \times 10^{-3}$ & $2.1$ & $4.08$ \\
		TRMC (RS, $\alpha$=50\%) & $2.616 \times 10^{-3}$ & $0.020 \times 10^{-3}$ & $3.2$ & $6.09$ \\
		FS                       & $2.504 \times 10^{-3}$ & $0.050 \times 10^{-3}$ & $1.0$ & $1.0$ \\ \hline 
	\end{tabular}    
	\label{tab:resampling}
\end{table}

\begin{figure}[tbp]
	\centering
	\includegraphics[width=80mm]{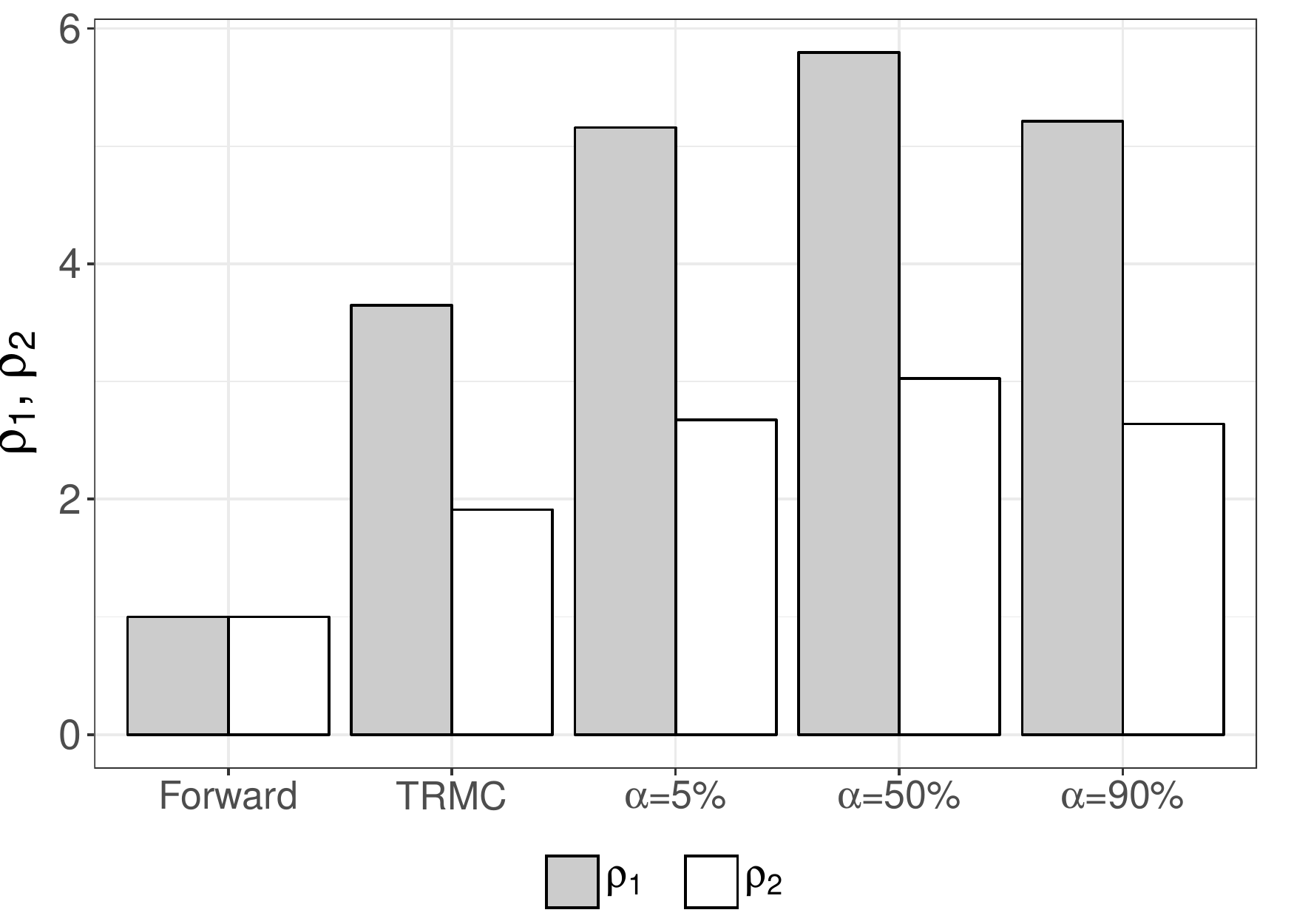}
	\caption{Comparison among FS (Forward), TRMC, and TRMC(RS) for the Lorenz 96 model.
		$\alpha = \alpha_{0}\%$ means TRMC(RS) with $\alpha = \frac{\alpha_0}{100}$.
	}
	\label{fig.lorenz96_rho_comparison}
\end{figure}

On the other hand, Fig.~\ref{fig.lorenz96_rho_comparison} shows that TRMC (RS) is more efficient than TRMC in a wide range of threshold values. We also show the weight distributions of TRMC and TRMC (RS) 
in Fig.~\ref{fig.resampling_weight_lorenz96}.
The variance of the distribution is much smaller for TRMC (RS) than for TRMC.

\section{Discussion}
\label{sec:dis}

The examples provided in the preceding sections show that backward simulations using TRMC provide correct averages and can be more efficient than forward simulations. In these examples, the computational efficiencies of TRMC are 3--16 times higher than those obtained by forward simulation, when the calculated probability of hitting the target is $2 \times 10^{-3}$--$10^{-5}$. 
Note that TRMC can be used to calculate the probability for an arbitrarily small target region; this would be impossible by using forward simulation.

There are, however, cases in which TRMC is inefficient. First, TRMC is not advantageous if the time-reversed paths rarely encounter a region in which the initial density $p(x_0)$ is high; this can occur when the initial density is not broad. Another case in which TRMC can be inefficient is when the weight in Eq.~\eqref{eq.ww_iba} (or, in the continuous time version, Eq.~\eqref{eq.factor_limit_2}) is highly time dependent. If paths with smaller weights in the initial stage of backward simulation acquire larger weights in the latter stage, resampling of the path (particle splitting) in SMC may not be effective. In this case, if TRMC with SIS is ineffective, TRMC with SMC also shows poor performance. 

To discuss the possible improvement of the algorithm, it is useful to introduce optimal backward dynamics. Although it is not easy to obtain these dynamics {\it a priori}, the formal definition is derived as follows. First, the marginal probability at step~$n$ obtained from the forward simulation is defined as
\begin{align}
	p(x_n) =\int dx_{0:n-1} \left\{ \prod_{i=0}^{n-1}  p\left( x_{i+1} | x_{i} \right) \right\} p(x_{0}) {}_,
	\label{eq:iba_mar1}
\end{align}
which satisfies the relation
\begin{align}
	p(x_{i+1}) =\int  p(x_{i+1}|x_i)p(x_i) dx_i {}_.
	\label{eq:iba_mar2}
\end{align}
Using Eqs.~\eqref{eq:iba_mar1} and \eqref {eq:iba_mar2}, the transition probability $q^*$ of the optimal backward dynamics is defined as
\begin{align}
	q^*(x_{i+1} \rightarrow x_i)= 
	\frac{p(x_{i+1}|x_i)p(x_i)}{\int  p(x_{i+1}|x_i)p(x_i) dx_i}
	=  \frac{p(x_{i+1}|x_i)p(x_i)}{p(x_{i+1})} {}_.
	\label{eq:iba_q}.
\end{align}
Note that Eq.~\eqref{eq:iba_q} appears similar to the formulas used in Bayesian inference when the probability $p(x_i)$ obtained by forward simulations is regarded as an analog of the prior distribution of $x_i$. In terms of the selection of Eq.~\eqref{eq:iba_q} for backward dynamics, the following relation holds:
\begin{align}
	\left\{ \prod_{i=0}^{n-1}  p\left( x_{i+1} | x_{i} \right) \right\} p(x_{0})
	=
	p(x_N) \prod_{i=N-1}^0  q^*(x_{i+1} \rightarrow  x_i),
	\label{eq:iba_path}
\end{align}
Eq.~\eqref{eq:iba_path}  means that the combined probability of time-reversed paths defined by forward simulation is recovered by the backward dynamics Eq.~\eqref{eq:iba_q}. Specifically, the time-reversed paths initialized by $p(x_N)$ automatically converge to their initial density $p(x_0)$ using the backward dynamics Eq.~\eqref{eq:iba_q}.
In this sense, $ q^*(x_{i+1} \rightarrow  x_i)$ in Eq.~\eqref{eq:iba_q} is considered as the optimal backward dynamics.
Implementation of these dynamics, however, requires the probabilities $p(x_i), i=1, \ldots N$, which are usually not available prior to the simulations. Note that the backward dynamics defined by the Langevin equation in previous studies~\cite{anderson1972factors, haussmann1986time, millet1989integration} can be derived from Eq.~\eqref{eq:iba_q} as a continuous-time limit.

Equations \eqref{eq:iba_q}  and  \eqref{eq:iba_path} were previously discussed~\cite{briers2010smoothing, lindsten2013backward, isard1998smoothing} in the field of time-series data analysis, where approximations of the marginal probabilities  $p(x_i), i=1, \ldots N$  are used to define the backward dynamics  $q(x_{i+1} \rightarrow  x_i)$. In these studies, the observed data were available at many of the time steps $i=0, \ldots, N$, whereas the target is given only at $i=N$ in our problem. Then, approximations of probabilities $p(x_i), i=1, \ldots N$ are derived using forward simulations constrained with the observed data (``filtering stage'').

It is, however, difficult to apply these methods to our problem. If we were to apply a similar method to our problem, we would have to run a number of forward simulations to estimate $p(x_i), i=1, \ldots N$ before executing backward simulations. This would be computationally expensive and seems unrealistic without a highly efficient method for the probability estimation. Methods such as those discussed previously~\cite{kappen2016adaptive, heng2017controlled} may be applied to optimize the backward dynamics in our problem, but this is left for future study.

On the other hand, when some observed data are available outside the equations that describe the stochastic process, we may use these data to approximate $p(x_i), i=1, \ldots N$ and hence use them to approximate the optimal backward dynamics. 
In this case, we avoid the use of a large amount of forward computation to construct the optimized backward dynamics. 
This seems possible for the stochastic typhoon model, where data from actual observations of real typhoons are available. 
Note that this idea is different from data assimilation (i.e., inference with simulations combined with observed data), because here we use observed data only to improve the computational efficiency; they do not cause the bias of calculated probabilities.   

In this paper, we assumed that the number of time steps is fixed.
This is not, however, always clear in advance in realistic problems.
In the case of a realistic typhoon model, we must consider the case of passing through Tokyo between two discrete time steps.
The interpolation method in this case will be developed in a future work.

\section{Concluding Remarks}
\label{sec:cr}

We discussed methods for the backward simulation of the stochastic process. These methods trace a time-reversed path from the target region to the initial configuration. A na\"{i}ve approach to this problem was shown not to function as expected. To resolve the difficulties, the time reverse Monte Carlo method (TRMC) was introduced. The TRMC method is based on SIS and SMC, and is designed to provide the probabilities of events correctly. TRMC with SIS was tested for the stochastic typhoon model and the Lorenz 96 model; it converges more efficiently than forward simulations in some of these examples. For simulations with a larger number of steps, TRMC with SMC was shown to be advantageous. We also discussed the limitation and possible improvement of TRMC and its relation to the Bayes formula.

\appendix

\section{Deviation of Eq.~\eqref{eq.factor_limit}}
\label{sec:appendix_a}
The aim of this appendix is to prove Eq.~\eqref{eq.factor_limit}.
Up to the first-order $\Delta t$, the Jacobian $\mathrm{det}(J_{g}(x))$ is given by
\begin{align}
	\mathrm{det}(J_{g}(x))
	&=
	\mathrm{det}
	\left(
	I
	+
	\nabla f(x) \Delta t
	\right) \nonumber \\
	&=
	1 
	+ 
	\Tr 
	\left(
	\nabla f(x) \Delta t
	\right)
	+ 
	O( \left(\Delta t \right)^2 ) \nonumber \\
	&=
	\exp \left[
	\mathrm{div} f(x) \Delta t
	\right]
	+ O( \left(\Delta t \right)^2 ),
	\label{appendb_expdiv}
\end{align}
where $I$ is the unit matrix of order $D \times D$.
$D$ is the dimension of stochastic process $X$.

Using Eq.~\eqref{appendb_expdiv},
we obtain in the limit as $\Delta t \rightarrow 0$
\begin{align}
	J(y_1, \dots, y_N)
	&=
	\prod_{i=0}^{N-1}
	\left|
	\mathrm{det}(J_{g^{-1}}(y_{i+1}))
	\right| \nonumber \\
	&=
	\exp \left[
	\sum_{i=1}^{N}
	- 
	\mathrm{div} f(x_i) \Delta t
	\right] 
	+
	O( \left(\Delta t \right)^2 )
	\nonumber \\
	& \xrightarrow[\Delta t \to 0]{} 
	\exp
	\left[
	-
	\int_{0}^{T}
	\mathrm{div}{f(x_t)} dt
	\right].
	\label{eq:jacobian_limit}
\end{align}
The above equation is Eq.~\eqref{eq.factor_limit} in the main text.

\section{Deviation of Eq.~\eqref{eq.factor_limit_2}}
\label{sec:appendix_b}
The aim of this appendix is to prove \eqref{eq.factor_limit_2}.

Up to the first-order $\Delta t$, the weight at time $t_i$ is given by

\begin{widetext}
	\begin{align}
		\label{eq:weight_limit_proof}
		W_{i}
		&=
		\exp
		\left[
		-\left( f(x_{i+1}) - f(x_{i}) \right)^{T}
		\Sigma^{-1}
		\left(
		(x_{i+1} - x_{i})
		-
		\frac{\Delta t}{2}
		\left(
		f(x_{i+1}) + f(x_{i})
		\right)
		\right)
		\right] 
		\\ \nonumber
		&=
		\exp
		\left[
		\Tr
		\left(  
		-\left( f(x_{i+1}) - f(x_{i}) \right)^{T}
		\Sigma^{-1}
		\left(
		(x_{i+1} - x_{i})
		-
		\frac{\Delta t}{2}
		\left(
		f(x_{i+1}) + f(x_{i})
		\right)
		\right)
		\right)
		\right] 
		\\ \nonumber
		&=
		\exp
		\left[
		-
		\Tr
		\left(
		\left( 
		\nabla f(x_i) 
		(x_{i+1} - x_{i})
		\right)^{T}
		\Sigma^{-1}
		(x_{i+1} - x_{i})
		\right)
		+
		o(\Delta t)
		\right] 
		\\ \nonumber
		&=
		\exp
		\left[
		-
		\Tr
		\left(
		\nabla f(x_i) ^{T}
		\Sigma^{-1}
		(x_{i+1} - x_{i})
		(x_{i+1} - x_{i})^{T}
		\right)
		+
		o(\Delta t)
		\right].
	\end{align}
\end{widetext}

In the limit as $\Delta t \rightarrow 0$, 
Eq.~\eqref{eq:fd} becomes the following stochastic differential equation:
\begin{equation}
dX_{t} = f\left(X_{t} \right) dt + \sigma dW_t, 
\label{eq:sde_limit}
\end{equation}
where $W_t$ is a standard Brownian motion.
Here, we used Ito's rule~\cite{joshi2003concepts, oksendal2013stochastic}, in which we substitute $\sqrt{dt}$ for $dW_{t}$ and consider up to the order of $dt$.  
Using Eq.~\eqref{eq:sde_limit}, we obtain the following relation in the limit as $\Delta t \rightarrow 0$
\begin{widetext}
	\begin{align}
		(x_{i+1} - x_{i}) (x_{i+1} - x_{i})^{T} 
		& \xrightarrow[\Delta t \to 0]{} 
		dx_{t} dx_{t}^{T}
		= 
		\left(f\left(x_{t} \right) dt + \sigma dW_t\right)
		\left(f\left(x_{t} \right) dt + \sigma dW_t\right)^{T}
		\\
		& =
		\sigma dW_t dW_t^T \sigma^Tdt
		+
		O(dt)			
		=
		\Sigma dt,
		\label{eq:dxdx}
	\end{align}
\end{widetext}
where we used the relationships $dW_t dW_t^T = dt$ and $\sigma \sigma^T = \Sigma$.

As a result, we obtain using Eq.~\eqref{eq:weight_limit_proof} and \eqref{eq:dxdx}
\begin{widetext}
	\begin{align}
		\prod_{i=0}^{N-1}W_i
		&=
		\exp
		\left[
		-
		\sum_{i=0}^{N-1}
		\Tr
		\left(
		\nabla f(x_i) ^{T}
		\Sigma^{-1}
		(x_{i+1} - x_{i})
		(x_{i+1} - x_{i})^{T}
		\right)
		+
		o(\Delta t)
		\right] \\
		&\xrightarrow[\Delta t \to 0]{} 
		\exp
		\left[
		-
		\int_{0}^{T}
		\Tr
		\left(
		\nabla f(x_t) ^{T}
		\right)
		dt
		\right] 
		=
		\exp
		\left[
		-
		\int_{0}^{T}
		\mathrm{div}{f(x_t)} dt
		\right],
	\end{align}
\end{widetext}
which is Eq.~\eqref{eq.factor_limit_2} in the main text.

\bibliographystyle{jpsj} 
\bibliography{ref}
\end{document}